\documentclass[aps,twocolumn,showpacs]{revtex4}
\usepackage{epsfig}%
\usepackage{dcolumn}
\usepackage{bm}
\begin{document}%
\title{\Large\bf 
Comment on ``Understanding the $\mu$SR spectra of MnSi without magnetic polarons"
}%
\author{Vyacheslav G.~Storchak}
\email{mussr@triumf.ca}
\affiliation{National Research Centre ``Kurchatov Institute'',
 Kurchatov Sq.~1, Moscow 123182, Russia}
\author{Andrey M.~Tokmachev}
\affiliation{National Research Centre ``Kurchatov Institute'',
 Kurchatov Sq.~1, Moscow 123182, Russia}

\vfil 
\date{9 September 2014} 
\begin{abstract}
Amato {\it et al.} have reported transverse field muon spin rotation experiments
performed on single crystal of MnSi in a single magnetic field of 5200~Oe 
at a single temperature of 50~K. 
They present the angular dependence of the muon precession frequencies
which they interpret in terms of dipolar magnetic field experienced by bare muons.
Such interpretation comes from a rather mechanistic approach 
without plausible physical backing: 
the wealth of experimental data collected so far 
does not justify this oversimplification.
No consideration is given to a fundamental feature of MnSi ---
strong magnetic field inhomogeneities on the scale of a lattice spacing 
found by many different techniques.
The computational procedure also raises a number of objections, 
in particular, applicability of Kohn-Sham 
DFT to strongly correlated systems like MnSi.
We demonstrate that the conclusion of ``Understanding the $\mu$SR spectra
of MnSi without magnetic polarons" is premature.  
\end{abstract}
\vfil
\pacs{71.38.Ht, 71.70.Gm, 72.80.Ga, 76.75.+i}
\maketitle

Recently, new muon spin rotation ($\mu $SR) measurements 
and associated calculations
have been reported 
in the strongly correlated electron (SCE) system MnSi
\cite{Amato2014}. 
The authors of Ref.~\cite{Amato2014} claim 
that their work provides understanding of 
$\mu $SR spectra of 
MnSi without invoking magnetic polarons. 
Indeed, there is a controversy concerning the 
origin of several frequencies in the $\mu $SR
spectra. The conventional approach is to 
ascribe them to magnetically inequivalent 
muon stopping sites. 
It typically assumes that muons stay bare. 
An alternative reasoning is 
based on high transverse field $\mu SR$ 
experiments demonstrating two-frequency spectra 
for the wide temperature range 2-305~K and 
magnetic fields up to 7~T. 
The spectra are explained 
by formation of spin (magnetic) polaron (SP) 
\cite{Storchak2011} --- a few-body state 
formed by a localized electron mediating ferromagnetic 
interaction between magnetic ions in its 
immediate environment. 

Amato {\it et al.} state that such electron localization 
is impossible in metals, referring to
a coupled $\mu^+$e$^-$ system (muonium, Mu)
which does not show up in metals
because of screening of Coulomb interactions 
\cite{Amato2014}.
However, formation of SP 
is an established phenomenon which is widely discussed in 
SCE metals 
(see {\it e.g.} \cite{Mott1990,Mott1993,Storchak2010}).
A possible reason for the misleading interpretation of SP as Mu is that 
both bound states are characterized by certain hyperfine couplings. 
However, one has to realize the fundamental difference between 
Mu atom and spin polaron.
Muonium
is formed in non-magnetic semiconductors and insulators 
entirely due to the Coulomb interaction between the $\mu^+$ and the $e^-$,
characterized by a two-line $\mu $SR 
spin precession spectrum \cite{Brewer1994} reflecting the 
electron-muon hyperfine interaction \cite{Patterson1988}.
In contrast,  
the basic interaction which causes SP formation is the exchange 
interaction ($J\sim 1$~eV) 
between a free carrier and local spins of the magnetic ions of the 
host, inducing 
electron localization into a ferromagnetic (FM) ``droplet'' 
on the scale of the lattice spacing \cite{deGennes1960}.
Thermodynamic responses of Mu and SP bound states --- 
dependences on magnetic field and temperature --- are also 
fundamentally different.
Mu {\sl cannot\/} be observed when its electron wavefunction 
overlaps with paramagnetic moments:
the strong pair exchange interaction 
of the bound electron with the host's spins 
(so-called ``spin exchange'' \cite{Patterson1988}) 
would result in extremely rapid spin fluctuations 
averaging the hyperfine interaction to zero.
In contrast, 
local ferromagnetic ordering 
mediated by the trapped SP electron 
holds the spin ``fixed'' and 
results in a non-zero hyperfine interaction.
The mere observation of hyperfine-split lines 
in the $\mu $SR spectra in a metal or in a 
magnetic material is a strong evidence for SP formation.


Here one has to make a clear difference between SP and 
bound spin polaron (BSP) typically detected 
in magnetic semiconductors, in particular, 
{\sl dilute\/} magnetic semiconductors (DMS, systems like CdMnTe or CdMnSe).
The formalism of DMS in terms of BSP
is justified by the low concentration of magnetic ions
which makes the exchange interaction small compared to the 
Coulomb interaction.
In SCE metals, in particular MnSi, 
magnetic ions are present in high concentration in every unit cell, 
which justifies the dominant role of the exchange interaction.   
Moreover, the current case deals 
with {\sl free, mobile\/} SP \cite{Mott1990,Mott1993}, 
as opposed to BSP in DMS.
The clear difference between free SP and BSP is 
that the former is spin-saturated while the latter is not.  
This difference is a direct consequence of 
the exchange-driven nature of the former 
versus the Coulomb-dominated origin of the latter.  
The BSP model 
(a large species of
the scale of many lattice constants) 
is hardly relevant to the current case
as the former is based on the concepts of 
(light) electron effective mass and (high) dielectric constant, 
both of which break down in the case of the SP in MnSi --- 
a {\sl small\/} species with a characteristic radius 
on the order of a lattice constant which  
exhibits strong mass {\sl enhancement\/} reported by various experiments 
on strongly correlated electron materials including MnSi.
  

In SCE metals SP do form (unlike Mu and BSP) and are detected :
as is the case for the more common lattice polaron, 
formation of a spin polaron 
may profoundly renormalize 
the bare electron band (bandwidth $\Delta_0 \sim 1-10$~eV) into a
much narrower ($\Delta_{\rm SP}/k_{\rm B} \sim 10^{-4}-10^{-3}$~eV) 
spin polaron band \cite{Mott1990,Nagaev2002}.
Such SP band supports coherent SP dynamics \cite{Mott1990}.
The band nature of the SP observed is evidenced by the equal amplitudes 
of SP lines in $\mu $SR spectra over the entire temperature 
range of their existence 
\cite{Storchak2010,Storchak2011,Storchak2012}.
As the exchange term in MnSi 
is the dominant interaction leading to SP formation 
(the Coulomb interaction is effectively screened), 
the role of the muon is reduced to that of a ``trapping center": 
the host lattice is populated by free (mobile) SP \cite{Mott1990}, 
one of which is captured by the muon, 
from which we detect two-frequency Mu-like spectra 
characteristic of a bound electron state, 
as in another SCE metal Cd$_2$Re$_2$O$_7$ \cite{Storchak2010}.

This scenario of muon-captured SP necessarily requires 
pure enough samples with long enough SP mean free pass 
so that residual defects and impurities do not affect 
the capture similar to electron capture by the muon 
in pure insulators and semiconductors 
to form Mu atom \cite{Storchak1995,Storchak1997,Storchak2004,Eshchenko2009}.
If the capture cross section for such defects and/or impurities
is higher than that for the muon, they prevent SP capture by the muon, 
which in this case stays bare. This is similar to what is observed 
in impure insulators and semiconductors which exhibit no or severely reduced 
Mu formation probability \cite{Eshchenko2002,Storchak2004}. 
Unfortunately, residual resistivity of MnSi sample used in Ref.~\cite{Amato2014} 
is not reported
to compare with that reported in Ref.~\cite{Storchak2011}.
Therefore we do not exclude such a possibility that pure enough samples 
studied in Ref.~\cite{Storchak2011} reveal SP spectra 
while the sample measured in Ref.~\cite{Amato2014} exhibits bare muon state.

In any case, the authors of Ref.~\cite{Amato2014} do not 
discuss any physical reasons that can 
differentiate the two interpretations. 
Instead, they provide a rather mechanistic approach 
where major features of the spectra are 
attributed to some local magnetic fields 
without much regard for their physical 
origin. The attempt to interpret the spectra 
should be commended but i) the claimed 
{\it understanding} can be reached only 
when the performed mathematical manipulations 
correspond to a consistent physical picture; 
ii) the methodology used raises more 
questions than provides answers.

First of all, it is necessary to reiterate 
the major physical reasons behind the spin polaron 
model of MnSi. When a coupled $\mu ^+e^-$ 
SP state is formed, muon spin-flip 
transitions produce a characteristic 
two-line spectrum in high transverse magnetic field 
with the splitting being 
determined by the muon-electron hyperfine 
interaction \cite{Storchak2009}. 
The dependence of the splitting on both magnetic 
field and temperature in MnSi corresponds well to 
the SP model \cite{Storchak2009} 
with the spin $S$=24$\pm $2 and the Bohr 
radius of electron $R\approx $0.4~nm. 
This $R$ sets up the characteristic length scale of the problem ---
a lattice spacing --- 
also found in several other SCE materials 
\cite{Storchak2009a,Storchak2009b,Storchak2010,Storchak2010a,Storchak2010b,Storchak2011a}
The fact that there is a characteristic 
dependence of the spectrum on the 
temperature and the field is totally 
ignored by the authors of Ref.~\cite{Amato2014}. 

In the paramagnetic region, the signal splitting 
is proportional to the bulk susceptibility. Such 
behaviour is expected for both spin polaron and 
bare muon models. In contrast, low-temperature 
studies can discriminate between the two models.
$\mu $SR experiments at 25~K \cite{Storchak2011} 
show that the splitting increases almost 2 times 
when the external magnetic field increases from 
0.5~T to 5~T (at this temperature all the data 
correspond to the ferromagnetically aligned 
MnSi) while magnetization increases 20 \% at 
most \cite{Demishev2012}, which is difficult to 
explain within the bare muon model which ascribes the 
splitting to the dipolar field induced by local 
Mn ions.

Furthermore, the two-line splitting is 
observed {\it at the room temperature} in the 
paramagnetic phase well above 
$T_C\approx $30~K, where fast spin 
fluctuations reduce any local fields at the bare 
muon to an average Knight shift from 
conduction electrons, which is 
typically 2 to 3 orders of magnitude less 
than the observed splittings \cite{Storchak2009,Storchak2010}. 
More fundamentally, the conventional explanation based on multiple 
bare muon sites should lead to abrupt change of 
the splitting at $T_C$ which does not show up in our 
experiment for a wide range of external magnetic fields \cite{Storchak2011}. 
This means that muons do not stay bare in MnSi 
and, therefore, do not act as local magnetometers. 
Instead, the fact that the line splitting does not exhibit 
a dramatic change at $T_C$ indicates that 
the local environment around the muon 
is fundamentally different from the rest of the host, 
which is consistent with local FM phase within SP similar to what is found 
in other magnetic materials \cite{Storchak2009b,Storchak2014}.
This experimental fact alone is capable to dismiss 
the entire picture of a bare muon in MnSi suggested in \cite{Amato2014}.
Furthermore, the deviations from the weak itinerant-electron 
magnetism model as revealed by electron spin 
resonance studies of MnSi are attributed 
to spin polarons \cite{Glushkov2011}.
A mid-infrared feature 
in optical conductivity spectra of MnSi
is an established fingerprint of a polaron species \cite{Mena2003}. 
Likewise, observation of a non-Fermi-liquid behavior at low temperature
and strong electron scattering cross sections reflecting 
inhomogeneities on a scale of the order of the lattice spacing
above 200~K is consistent with SP.
Finally and most fundamentally, 
both microscopic magnetic field inhomogeneities on the scale of the lattice spacing
discovered by neutron scattering \cite{Pfleiderer2004}, NMR \cite{Yu2004} 
and $\mu$SR \cite{Uemura2007}, and 
an effective-mass enhancement are also consistent 
with the lattice-spacing-size SP formation.

The conclusion of Ref.~\cite{Amato2014} 
about the multiple muon sites is based on a 
series of fits of $\mu SR$ signals by four 
components with equal amplitudes. The authors 
attribute them to structurally equivalent muon 
sites (4a Wyckoff position). To prove this 
hypothesis they consider the weak angular 
dependence of the frequencies with respect 
to rotation of the sample. The 
symmetry of the 4a Wyckoff 
position means that the sum of dipolar 
contributions for the 4 components should 
be exactly zero irrespective of the 
sample rotation angle:
\begin{equation}
\sum _{i=1}^{4}B_{dip,i}(\phi )=0.
\end{equation}
This condition is not satisfied for the 
angular-dependent parts of the fitted 
frequencies and the sum gives some 
residual field $B_{res}(\phi )$. This field is 
relatively large with the amplitude close to that 
of one of the fitted signals. Instead of considering 
the source of the discrepancy the authors of 
Ref.~\cite{Amato2014} {\it arbitrarily} modify 
all the signal frequencies by subtracting 
$B_{res}(\phi )/4$ functions from each of them 
and calling this correction demagnetization 
field. It means that the experimental data are 
put into Procrustean bed of the symmetry of 
4a Wyckoff position for {\it all} values of the 
sample rotation angle.

The angular part of each signal is then 
fitted by 3 parameters of the dipolar tensor 
in the reference frame. The authors of 
Ref.~\cite{Amato2014} claim that each signal 
provides us with the full set of parameters, 
namely, the parameter $a_{dip}$ representing 
dipolar tensors in the reference frame of 
the crystal and Euler angles $\theta $ and 
$\phi $ corresponding to orientation of the 
sample with respect to crystallographic axes. 
This is not true. First, the fitted 
tensor components are not independent because 
the modification of the angular dependencies 
(see above) artificially forced them to 
satisfy the symmetry conditions:
\begin{equation}
\sum _{i=1}^{4}A_{kl}^{i}=0.
\end{equation}
Second, it is common knowledge that an 
arbitrary rotation of a solid is given by 
three (not two) Euler angles. Two angles 
define only the plane of rotation, while 
the third angle defines the orthogonal 
axes in the plane (or, alternatively, 
the direction corresponding to zero 
rotation angle). It is impossible to 
find 4 independent parameters from 3 
dipolar tensor components $A_{xx}$, 
$A_{yy}$ and $A_{xy}$. Remarkably, the 
authors of Ref.~\cite{Amato2014} have 
somehow chosen the zero rotation angle 
direction of the MnSi sample in their 
experiments corresponding exactly to the 
best fit of the signals by only two 
Euler angles. This preknowledge of the 
third Euler angle certainly needs 
explanation and makes the whole procedure 
very questionable. Moreover, our analysis 
shows that the quality of the fits is not 
that great to estimate parameters with 
such high precision (like $a_{dip}$ 
determined to be -0.2044(40)~mole/emu).

The angular dependence of the fitted 
frequencies (quite arbitrarily divided 
into dipolar and demagnetization field 
contributions) explains only a small 
part of the deviation of signal from 
the free muon frequency. This large 
negative shift is explained in 
Ref.~\cite{Amato2014} by the contact field 
arising due to spin-polarization of 
the conduction electrons at the muon 
site. At this high temperature such a large 
contribution cannot be due to 
the Knight shift. Again, there is a 
question of precision: the authors of 
Ref.~\cite{Amato2014} claim that in 
equation 
\begin{equation}
{\bf B}_{cont}=A_{cont}{\bf \chi B}_{ext}
\end{equation}
$B_{cont}$ is determined with 
precision 4\%, $\chi $=0.030~emu/mole, 
but they find $A_{cont}$ from these 
data with much higher precision 
(-0.9276(20)~mole/emu).

Even if one accepts the interpretation based 
on the huge hyperfine contact coupling tensor 
due to conduction electrons, 
its transferability between 50 K 
(paramagnetic phase) and 5 K (helimagnetic 
phase), defining the analysis of zero-field 
spectra in Ref.~\cite{Amato2014}, is doubtful.
It is established that the density of states 
near the Fermi level is quite different for 
magnetically ordered and paramagnetic MnSi 
\cite{Jarlborg2007}. Therefore, the assumption 
that "no massive changes occur on the Fermi 
surface when crossing $T_C$" which is at the 
heart of zero-field (low-temperature) data 
discussion of Ref.~\cite{Amato2014} is at 
least questionable and needs substantiation. 
Without that any correspondence between the 
calculated and experimentally observed 
frequency can be rendered coincidental.
Another highly questionable approximation 
in the analysis of the ZF-$\mu $SR spectrum 
is that the local magnetization on the muon in 
Eq. (13) is assumed to be an equally 
weighted sum of Mn moments within a sphere 
of one lattice constant radius and neglecting 
all the rest --- definitely not the distance 
dependence expected for the RKKY interaction.

To support their findings the authors of 
Ref.~\cite{Amato2014} performed a 
quantum-mechanical calculation. Its purpose 
is not clear. First, 
the calculation was performed for the 
ferromagnetic state of MnSi, while all 
the experiments were made for 
helimagnetic and paramagnetic phases. 
Second, the method of calculation is 
inappropriate for MnSi. Density 
functional theory (DFT) is a quantum 
computational method replacing the 
solution of Schroedinger equation by 
minimizing an energy functional of 
one-electron density. The exact form 
of the functional is not known and 
different approximations for 
exchange-correlation part of the density 
functional are used: local density 
approximation stemming from the known 
solution for the homogeneous electron 
gas as well as GGA and meta-GGA corrections 
trying to improve the functional by 
adding contributions from the first and 
the second derivatives of the density.
In practice, standard DFT uses Kohn-Sham 
approximation where the density matrix is 
defined as coming from a set of 
non-interacting one-electron 
quasiparticles, which corresponds to 
the one-electron band picture in the case 
of periodic systems. DFT is known to 
be a working horse for electronic 
structure calculations of solids. It 
provides reliable results for many 
systems but it is far from being 
universal. The deficiencies 
of the standard (based on the 
Kohn-Sham approximation) 
DFT method are well publicized and 
explained on the examples of 
characteristic failures like 
potential energy curves of H$_2^+$ 
(delocalization error) and H$_2$ 
(static correlation error) 
\cite{Cohen2012}. As a 
consequence, there are general 
guidelines defining the classes of 
problems where the results should 
be considered with a grain of salt 
(like weak interactions) and where 
the method should not be applied 
at all (like studies of global 
potential energy surfaces and 
systems with strong electron correlations). 

MnSi is known to be strongly 
correlated and exhibiting non-Fermi 
behaviour, {\it i.e.} it is exactly 
a system where the standard (Kohn-Sham) DFT 
is expected to fail and it indeed 
fails. DFT calculations of different 
flavours systematically predict the
ground state of MnSi to have magnetic 
moment on Mn close to 1~$\mu _B$ 
\cite{Lerch1994,Jeong2004,Carbone2006,Collyer2008}, 
while the experimental value is 
0.4~$\mu _B$. A possible reason 
for this is known from XAS spectra 
\cite{Carbone2006}: MnSi has a 
mixed-valence ground state with 
significant on-site electron 
correlations. Therefore, quite 
expectedly, the authors of 
Ref.~\cite{Amato2014} has got the same 
wrong ground state with magnetic 
moment on Mn close to 1~$\mu _B$. 
The use of time-consuming full 
potential approaches and generalized 
gradient approximations is absolutely 
irrelevant to the problem because 
single reference calculations cannot 
describe systems with essentially 
non-idempotent density matrices. It is 
also worth noting that the approach 
used in Ref.~\cite{Amato2014} is 
not capable to find magnetic polarons 
in MnSi.

It is not clear why the authors 
do not provide details of 
calculations of muon embedded into 
MnSi. Surely, DFT calculations of 
muon stopping sites are routine and 
2$\times $2$\times $2 supercell 
calculations of MnSi with muon are 
not computationally too demanding. The 
electrostatic potential minima can 
be successfully used for 
determination of electrophilic 
attacking sites but their application 
for finding equilibrium muon sites is 
questionable (especially in 
combination with the wrong ground state) 
because muon is not a small charge probe 
and it can perturb its environment 
significantly. The mixed-valence 
character of the ground state indicates 
that the local magnetic structure can 
be also affected (although the correct 
description of this effect is beyond 
the capabilities of the standard DFT 
approach used in Ref.~\cite{Amato2014}).

In summary, the argumentation presented above
cast serious doubts upon conclusion of 
Ref.~\cite{Amato2014}) that $\mu$SR spectra of MnSi
can be correctly understood without invoking spin polarons.

\end{document}